# Graph-based hierarchical record clustering for unsupervised entity resolution

Islam Akef Ebeid
Department of Information Science
University of Arkansas at Little Rock
Little Rock, Arkansas, USA
iaebeid@ualr.edu

John R. Talburt
Department of Information Science
University of Arkansas at Little Rock
Little Rock, Arkansas, USA
jrtalburt@ualr.edu

Md Abdus Salam Siddique
Department of Information Science
University of Arkansas at Little Rock
Little Rock, Arkansas, USA
msiddique@ualr.edu

## Abstract

Here we study the problem of matched record clustering in unsupervised entity resolution. We build upon a state-of-the-art probabilistic framework named the Data Washing Machine (DWM). We introduce a graph-based hierarchical 2-step record clustering method (GDWM) that first identifies large, connected components or, as we call them, soft clusters in the matched record pairs using a graph-based transitive closure algorithm utilized in the DWM. That is followed by breaking down the discovered soft clusters into more precise entity clusters in a hierarchical manner using an adapted graph-based modularity optimization method. Our approach provides several advantages over the original implementation of the DWM, mainly a significant speed-up, increased precision, and overall increased F1 scores. We demonstrate the efficacy of our approach using experiments on multiple synthetic datasets. Our results also provide evidence of the utility of graph theory-based algorithms despite their sparsity in the literature on unsupervised entity resolution.

## 1 Introduction

Entity resolution is crucial in data cleaning, curation, and integration (Talburt et al., 2020). It also refers to finding duplicate records within the same table, across various tables, or multiple databases. Traditional and supervised approaches in entity resolution rely on handcrafting rules for matching records. However, the move towards automating entity resolution for data cleaning, curation, and integration has become the goal for many organizations. Thus, unsupervised entity resolution methods have proliferated, relying on an automated pipeline of preprocessing, blocking, matching, clustering, and canonicalization. However, unsupervised approaches suffer from higher inaccuracies than supervised ones due to inefficiencies in the used approaches, especially in the clustering step, where the goal is to discover latent unique entities.

The base of any entity resolution system is string similarity or fuzzy matching. Traditional and supervised entity resolution relies heavily on human input to guide the entity matching process using predefined rules. Defining those rules depends on handcrafting simple lexical, semantic, and syntactic conditions for matching records based on attribute similarity as in (Syed et al., 2012) and in the OYSTER system (Talburt & Zhou, 2013). In recent years unsupervised entity resolution has become crucial for scaling and generalizing the entity resolution process. Unsupervised entity resolution approaches could be divided into probabilistic methods, machine learning methods, and graph-based methods. Some work has been done in probabilistic unsupervised entity resolution that relies on estimating statistical models (Herzog et al., 2007; Lahiri & Larsen, 2005; Tancredi & Liseo, 2011). However, probabilistic methods that rely more on natural language processing, fuzzy matching, and token frequency have proliferated (Bilenko et al., 2003; Papadakis et al., 2011; X. Li et al., 2018; Alsarkhi & Talburt, 2018).

Probabilistic methods in unsupervised entity resolution typically rely on an automated pipeline of preprocessing, blocking, matching, clustering, and canonicalization. Preprocessing refers to multiple steps that involve merging and parsing data files, tokenizing, and normalizing the unstandardized references. Blocking refers to the strategy used to mitigate the quadratic complexity of pairwise comparisons in unsupervised entity resolution. That strategy relies on quick and dirty techniques that divide the preprocessed unstandardized references into chunks or blocks, avoiding string matching across the whole dataset. Each block can be processed separately, where pairwise string similarity can be applied with less computational cost (Papadakis et al., 2020). The matched pairs from each processed block are then combined and clustered to infer the unique latent entities represented in the data file. The clustering process aims to resolve any conflicts in pairwise matching and find records that indirectly match. Those conflicts typically occur due to the reliance on frequency-based blocking in unsupervised entity resolution systems (Talburt et al., 2020). Many challenges riddle the unsupervised entity resolution process; the most critical are parameter tuning, finding appropriate thresholds for matching probabilities, and exact record clustering. In addition, that process might need to be applied multiple times due to the low accuracy of relying on fuzzy matching alone without rules as in supervised methods.

### 1.1  Contribution

Here we study the problem of record clustering in probabilistic unsupervised entity resolution. We propose a graph-based 2-step hierarchical record clustering technique. We modify a state-of-the-art algorithm in unsupervised entity resolution named the Data Washing Machine (DWM) (Talburt et al., 2020). Integrating both techniques leverage the power and precision of graph theory methods and the speed and robustness of probabilistic token-based methods. Despite its advantages, the DWM presents some drawbacks and limitations. For example, the iterative approach used in the DWM relies on assessing the quality of the entity clusters using Shannon's entropy-based metric. The metric is difficult to interpret due to arbitrary threshold setting; hence it causes unnecessary iterations and higher computational costs.

More specifically, we modify the mechanism of iterating over a full, unsupervised entity resolution pipeline multiple times and replace it with a single shot graph-based approach that minimizes and localizes iterations to discover and optimize entity clusters. The optimization approach is done hierarchically and over two steps after modeling the matched record pairs as edges in a record graph. The first step relies on iteratively discovering large connected components or soft clusters in the unweighted, undirected record graph after applying a threshold to prune the record graph using a graph-based transitive closure algorithm named CC-MR (Kolb et al., 2014). The second step relies on breaking down those large soft clusters characterized by a high recall into more precise entity clusters also iteratively by adapting a graph-based clustering and community detection algorithm called Louvain's method (Blondel et al., 2008). The result is a significant speed-up when compared with the original implementation of the DWM. In addition to a significant increase in precision when applied to the synthetic benchmark data set described in (Talburt et al., 2009) Thus our contributions are summarized as:

- Applying Louvain's method (Blondel et al., 2008) to the unsupervised entity resolution problem
- Introducing Modularity Hierarchical  Record Graph Clustering as a hierarchical method that combines Louvain's method with a graph-based Transitive Closure algorithm named CC-MR (Kolb et al., 2014) used in the original DWM in a hierarchical 2-step clustering efficient approach
- Modifying the DWM framework to employ our graph-based hierarchical 2-step record clustering approach instead of the iterative self-assessing approach based on Shanon's entropy. Our modified graph-based framework will be referred to as GDWM throughout the paper.
- Contributing to the literature and methods in graph-based unsupervised entity resolution and showing that a hybrid method; combining both probabilistic unsupervised entity resolution methods with graph theory approaches, provides better results than using graph-based approaches alone (Zhang et al., 2020) or probabilistic token-based approaches alone (Talburt et al., 2020)
- Performing various experiments on a sampled synthetic benchmark dataset (Talburt et al., 2009) with known ground truths to demonstrate the effectiveness, accuracy, and robustness of our approach

The rest of the paper is organized as follows. In section 2, we describe previous work that has been done in graph-based approaches in entity resolution. In section 3, we describe both the algorithm and the method proposed. Section 4 describes our experiments on the proposed approach using a synthetic dataset in various configurations and settings. In section 5, we discuss the results and the implications of the proposed approach. Finally, we conclude in section 6. The full bibliography is available in section 7.

## 2 Related work

Here we review the literature on graph-based methods in entity resolution. Despite their sparsity in the literature on unsupervised entity resolution, graph-based methods and algorithms have been adapted before to the problem of entity resolution.

### 2.1 Token-based graph entity resolution

In token-based graph entity resolution, the goal is to construct a bipartite undirected graph of token nodes and record nodes and cluster the record nodes into unique entities using methods such as SimRank (Jeh & Widom, 2002). In (F. Wang et al., 2013), the authors introduced a graph-based entity resolution model. The model transformed the input data set into a graph of unique tokens where connectivity reflects the co-appearance of tokens in references. The graph was clustered using a weight-based algorithm that considered three types of vertices: exemplar, core, and support vertices. The algorithm then constructed r radius maximal subgraphs from the original token graph to discover clusters related to unique entities. Token-based methods, however, are computationally expensive and memory intensive due to the lack of an integrated blocking strategy.

### 2.2 Record-record similarity-based graph entity resolution

Record-record similarity graphs link structured unstandardized references in a weighted undirected graph where the nodes represent unique records. The connectivity represents the degree of similarity between individual references. That approach of constructing a record graph allows to directly utilize a whole set of graph clustering algorithms that graph theory and network science researchers have already developed, as suggested in (Hassanzadeh et al., 2009). While (H. Wang et al., 2016) applied a graph clustering algorithm to optimize minimal cliques in the graph. The algorithms approximated the NP-hard graph clique problem through pruning. Moreover, (Saeedi et al., 2018) developed the FAMER framework to combine multi-source data using blocking, matching, and clustering schemes. The framework modeled the merged data as a similarity-record graph and then leveraged graph clustering techniques to resolve the entities.

Other work has leveraged the graph's structure instead of just the weights between records. In (Draisbach et al., 2019), the authors proposed three algorithms to cluster the similarity graph based on structure rather than edge weights. They argue that graph-based transitive closure, such as in (Kolb et al., 2014), produces high recall but low precision because the graph's structure is not considered during clustering. They justified using maximal clique algorithms to leverage the graph's structure, which increases precision. There are also centrality and node importance-based methods where the edge weights are not considered, and node scores are propagated, such as in (Kang et al., 2020). In addition, the authors introduced the notion of a node resistance score in a co-authorship graph to model entity similarity. Node resistance can be thought of as a PageRank score (Page et al., 1999), where a random walker computes the probability of getting from the source node to the target node iteratively until convergence. Also (M. Sadiq et al., 2020), the authors introduced a graph-based model that linked two graph datasets by aggregating similarity scores from neighboring record nodes. However, record-record similarity methods are complicated and require extensive graph theory knowledge to tune the adapted methods.

### 2.3 Hybrid graph-based entity resolution

Hybrid methods that combine token-based bipartite graphs and similarity-based record graphs have been investigated in (Zhang et al., 2020) and (Zhang et al., 2018). The authors proposed an algorithm that combines text similarity with a graph-based algorithm. They first partition the data into a bipartite graph of record pair nodes and frequent term nodes to learn a similarity score of the record pair nodes. Then, the result was used to construct a record-record graph used to power the CliqueRank algorithm, which runs on the blocks of records identified by the first part, known as the ITER algorithm. The probability of a matching pair of records is then updated iteratively. The authors combined two distinct methods: the random walk-based approach and the graph clustering-based approach. However, the authors used the graph approach to match pairs of records without introducing any clustering approach that would resolve the entities.

The following section describes the framework, method, and algorithm used in GDWM.

### 3 Method (2000 words)

#### *3.1 Framework*

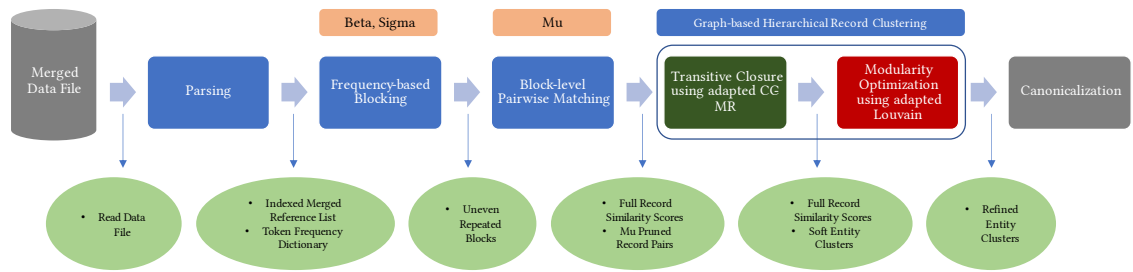

Figure 1. GDWM modifies DWM (Talburt et al., 2020) to eliminate the need for iterations and entropy-based evaluation by further breaking down initial soft clusters from the transitive closure algorithm described in (Kolb et al., 2014). To achieve that, we adopt a second clustering step based on Louvain's community detection method (Blondel et al., 2008).

#### *3.1.1 Merged Data File*

Merging data files is a crucial step in entity resolution. Data files and record sets usually come in multiple files containing unstandardized duplicate records. Those files are generally of different sizes and layouts, making it challenging for entity resolution algorithms. The merging of data files is usually the first step in any entity resolution framework. Sometimes using a single layout and sometimes combining without knowing the specific layout of the metadata. A standard layout means that the data is organized in the same fields. In our case, we merge all incoming data files into one file without relying on a specific layout where each row has a unique identifier, and the unique entity clusters are unknown. More formally, a record set $R$ consists of tuples $r_i = (u_i, f_i)$ where $u_i \in U$ is a unique identifier and $f_i \in F$ is a set of attributes with unknown headers.

#### *3.1.2 Parsing*

The parsing process includes preprocessing and tokenization and is central to our unsupervised entity resolution approach. Tokenization is preceded by a preprocessing step where the text from all the provided attributes and fields is combined into one string and normalized to the upper case. That is followed by removing special characters from the text. We use the standard approach to tokenizing text in English, splitting the text based on spaces between tokens. The parsed data are then loaded into memory, and a dictionary of unique token frequencies is computed. More formally, the processed data file contain a unique distinct set of tokens $T$ with different counts $C$ where $t_i$ is a distinct unique token where $\{t_i \in T\} \subseteq f_i$ and $c_i$ is the corresponding count of each unique token $t_i$ where $\{c_i \in C$ where $C = |\{t_i \in T\}|\}$.

#### *3.1.3 Blocking*

The blocking process is at the heart of our unsupervised entity resolution method. The goal is to reduce the potential number of string similarity operations across a data file as much as possible. That relies on inexpensive techniques to quickly filter out records in the data file with a very low probability of matching. Frequency-based blocking typically assumes that two records that refer to the same entity should share at least one token. Other blocking techniques include Locality-Sensitive Hashing and Random Sampling (Papadakis et al., 2020). Here we adapt the frequency-based blocking technique presented in the DWM (Talburt et al., 2020). The blocking method relies on two parameters, beta $\beta$ and sigma $\sigma$. First each reference $r_i \in R$ is processed using the sigma threshold to filter out tokens $t_i \in r_i$ with frequency $\{c_i \in C$ where $C = |\{t_i \in T\}|\}$ above sigma. That process could be thought of as a stop word removal where tokens that appear too much are considered non discriminative and unimportant. Following for each reference token $t_i \in r_i$ with frequency $\{c_i \in C$ where $C = |\{t_i \in T\}|\}$ below beta is considered a blocking token $\{t_{Bi} \in T\}$. Blocking tokens are identified for each reference in a list $L$ where the filtered records are repeated. The list is then grouped by blocking tokens regardless of reference. Each block will include all the references where the same blocking token appeared and the number of blocks will be equivalent to the number of unique blocking tokens in the dataset. This process is formalized in Algorithm 1.

---

**Algorithm 1:** Blocking

**Input :** $C(T)$ $unique$ $token$ $frequencies$, $R$ $record$ $set$, $\beta, \sigma$

**Result:** $B$ $list$ $of$ $blocks$

$B \leftarrow list$

**for** $r \in R$ **do**

    **if** $c_i \in C$ $for$ $each$ $t_{ij} \in r$ $where$ $t_i \in T \geq \sigma$ **then**

        $r \leftarrow remove$ $t_{ij}$ $from$ $r$

    **end**

    **if** $c_i \in C$ $for$ $each$ $t_{ij} \in r$ $where$ $t_i \in T \leq \beta$ **then**

        $L \leftarrow (t_{ij}, r)$

    **end**

**end**

$L_s \leftarrow sort(L, t_i \in T)$

$T_u \leftarrow unique(t_{ij} \in L_s)$

**for** $t_i \in T_u$ **do**

    **for** $l_s \in L_s$ **do**

        **if** $t_i = t_i \in l_s$ **then**

            $b_i \leftarrow r \in l_s$

            $B \leftarrow B + b_i$

        **end**

    **end**

**end**

**return** $B$

---

### 3.1.4 Pairwise Matching

After blocks have been identified, the next step is pairwise matching between references in each block. Even though any string similarity metric can be used on the blocked references to identify matches, we choose to use a string similarity algorithm named the Scoring Matrix used in (Talburt et al., 2020) and developed in (Li et al., 2018) and tested in (Al-Sarkhi & R Talburt, 2018) as an adapted Monge-Elkan (Monge et al., 1996) method. The Scoring Matrix method relies on lining up stop word filtered tokens of both references of the two records at hand as rows and columns of a matrix in order of appearance. The cell values of the matrix are the normalized Levenshtein's Edit Distance between each token in the first and the second records. After computing the edit distance between each token, the algorithm then normalizes the edit distance between all the rows and columns in the matrix that are non-zero dividing by the largest length between the two tokens. The algorithm then loops on all the rows and columns and averages all the largest normalized edit distances in the matrix for each row token or column token and that becomes the final normalized similarity score between the two references. We will not formalize the Scoring Matrix algorithm here as more details can be found in (Li et al., 2018).

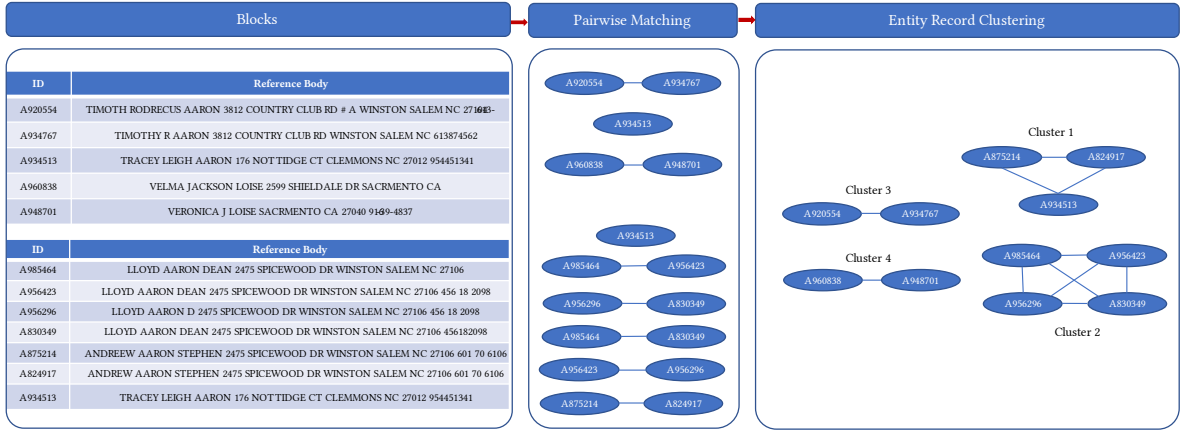

Figure 2. The typical unsupervised entity resolution approach starts with frequency-based blocking followed by pairwise matching, finally, entity clustering.

### 3.1.5 Graph-based Hierarchical Record Clustering (GDWM)

As mentioned before, entity resolution aims to find the unique latent entities in a dataset that multiple records might represent. The blocking algorithm considers two similar records representing the same entity if they have a minimum of one token in common. That increases the number of records representing one entity, as shown in figure 2. As a result, a record may appear in more than one entity, making the process of clustering more expensive and the underlying graph of matched records more complex. In addition, the matching process also implies that one record might be similar in an indirect way to another record by an algebraic property known as transitivity. In a simple sense, if record a matches record b and record b matches record c, then by transitivity record a matches record c. Hence transitive closure is a widely used approach to finding binary relations in ordered sets of pairs. When adapted to entity resolution, it also considers matched records as an edge list in a graph.

Hence clustering can be formalized by reformulating the problem as a graph $G = \{V, E\}$. A vertex/node $\{v \in V\}$ represents one record $\{r \in R\}$ where $\{r = v\}$ while an edge $\{e \in E \text{ where } E \subseteq V \times V\}$ represents whether two records $\{e_i = (r_i, r_j)\}$ are matched and an edge weight $w_e \in \{W(E): E \rightarrow \mathbb{R}\}$ represents the normalized similarity between the two records calculated in the pairwise matching step using the Scoring Matrix algorithm (Li et al., 2018). We also define a set of clusters $Q \subseteq P(V)$ where elements of Q are a subset of V and P is a partition of V. The clustered graph conventionally can be seen as a graph of subgraphs where each meta-vertex represents each subgraph of vertices or records. Hence this hierarchical relationship can be formalized as follows and also fits in our framework as illustrated in figure 3:

$$V' = Q$$

$$E' = \{(Q_i, Q_j): there\ exists\ v_i, v_j\ such\ that\ v_i \in Q_i, v_j \in Q_j, (v_i, v_j) \in E\}$$

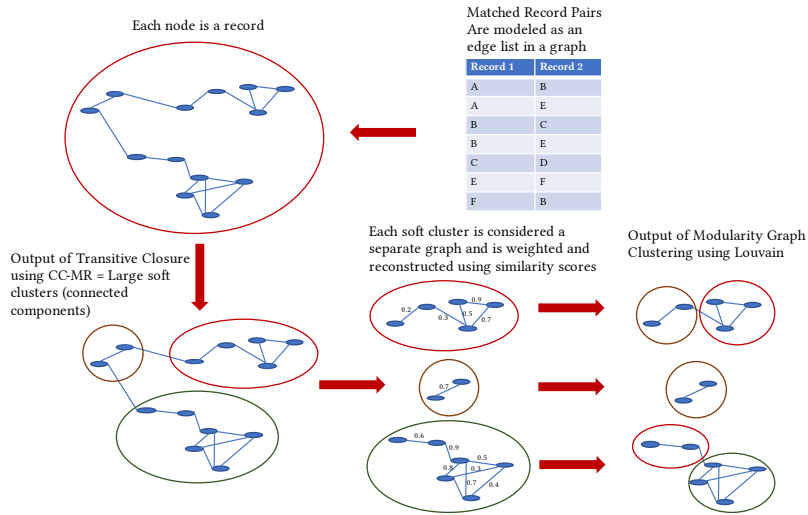

Figure 3. We view the record clustering step that follows pairwise matching as a graph clustering problem. The matched pairs with their similarity scores are considered an edge list of a graph. Then they are pruned using the threshold Mu. (Kolb et al., 2014) is applied to discover largely connected components or soft clusters. Each soft cluster is then remodeled as a complete separate undirected weighted graph and is pruned based on the original pair list and not mu. (Blondel et al., 2008) is applied to each graph, and final resolved entity clusters are achieved.

Here we view the post pairwise matching entity clustering process as a two-step process. First, we formulate the problem as a graph and apply a hierarchical graph clustering approach to discover the latent entity clusters. Second soft clusters are discovered using the efficient graph-based transitive closure algorithm utilized in the DWM (Talburt et al., 2020) and adapted from (Kolb et al., 2014) called CC-MR. The input to CC-MR is a pruned unweighted undirected graph of matched pairs. Next, we convert each soft cluster into a complete undirected weighted graph and prune it from the previously computed similarity scores between matched pairs. Finally, a Modularity optimization algorithm adapted from (Blondel et al., 2008) is used to hierarchically discover and refine the original soft clusters increasing the precision of the overall clustering process while eliminating the need for reiterations compared with the DWM framework. Our 2 step hierarchical method is described formally in Algorithm2.

**Algorithm 2:** Modularity Hierarchical Record Graph Clustering (GDWM)

**Input :** $S, \mu$ :
      *Pairs of matched records and their computed similarity scores, $Mu$*
**Result:** $Q$ : *list of clusters indexed by the least record node in each cluster*
$P \leftarrow TransitiveClosureUsingAdaptedCCMR(S, \mu)$
 *returns connected components as soft clusters see Algorithm 4*
$P_s \leftarrow sort(P, r \in P)$
$q_c \leftarrow unique(r \in P_s)$
**for** $c \in q_c$ **do**
    **for** $p(c) \in P_s$ **do**
        **for** $n1 \in p(c)$ **do**
            **for** $n2 \in p(c)$ **do**
                **if** $n1 \neq n2$ and $S(n1, n2)$ exists or $S(n2, n1)$ exists **then**
                    $E' \leftarrow append(n1, n2, s \in S)$
                **end**
            **end**
        **end**
    **end**
    $V' \leftarrow$ *initialize node clusters randomly*
    $V' \leftarrow AdaptedLouvain(G(V', E'))$ *see Algorithm 3*
    $V' \leftarrow$ *append single nodes as separate clusters*
    $M \leftarrow ComputeModularity(V', E')$
    **if** $M \geq 0$ **then**
        $Q \leftarrow$ *rename cluster ids to least record in cluster $V'$ and append*
    **end**
    **if** $M \leq 0$ **then**
        $Q \leftarrow$ *assign records in $V'$ to one cluster and append*
    **end**
**end**
**return** $Q$

### 3.1.5.1 Transitive Closure using adapted CC-MR

As mentioned before, the ordered list of matched records can be considered an edge list or an adjacency matrix representing a graph of records. Following a simple graph clustering approach to finding the strongly connected components in the graph could be used to perform transitive closure as the first step in our 2 step clustering approach. CC-MR was introduced in (Kolb et al., 2014) and used in (Talburt et al., 2020) as a single-step clustering method for the DWM starts by creating star subgraphs from the matched pair list. That is done by simply grouping the pair list by the smallest node. Then, the algorithm iteratively checks whether its vertices are assigned to subgraph components where the center of the subgraph component is the smallest vertex in its first-order neighborhood, relying on the MapReduce framework for scalability. The DWM provides an efficient implementation of the algorithm without the MapReduce approach. The algorithm is formalized and described thoroughly in (Kolb et al., 2014). Also, we provide a short description of the implementation we used in Algorithm 4.

```
Algorithm 4: Transitive Closure Using Adapted CC-MR
─────────────────────────────────────────────────────────────
Input : S, μ :
          Pairs of matched records and their computed similarity scores, Mu similarity threshold
Result: P  : List of soft clusters as pairs
          indexed by the least record in the cluster as the first element of the pair
P ← list of soft clusters
R_P ← initialize list of tuples
for s ∈ S do
    if s_i ≥ μ then
     |  R_P ← append s_i
    end
end
R_P ← sort by first element in pair
while no convergence do
    for (r_i, r_j) ∈ R_P do
        P ← append pair belonging to connected component
            when assuring that all record
            nodes belong to the connected component with the smallest record id
            node at the center in their neighborhood
    end
end
return P
```

### 3.1.5.2 Modularity Optimization using adapted Louvain

Next, we pass the soft clusters produced by the CC-MR transitive closure in addition to the previously computed similarities between records on the block level to our Modularity optimization algorithm adapted from the Louvain method described in (Blondel et al., 2008). In the original DWM implementation after transitive closure, an evaluation step is applied where each cluster is evaluated using Shannon's entropy-based approach. Then, a normalized quality metric is computed for each cluster, and a threshold is applied to decide whether the cluster shall be reiterated on or canonicalized. The clusters with qualities lower than the threshold are reiterated starting from blocking, where the similarity threshold parameter mu is incremented, and the lower quality clusters are further broken down. That process carries on until mu reaches one or no clusters are left below the quality threshold.

Multiple issues emerge with that approach. The most salient is trying to tune the similarity parameter mu, and the quality threshold epsilon is tricky, as shown in figure 5. In addition, clusters that are good enough but their quality have been slightly below the threshold might stay in the pipeline for 2 or 3 iterations increasing the computational cost unnecessarily. It is worth mentioning that these issues are addressed in a later version of the DWM using non-graph-based approaches. Nevertheless, in the current version that we are working with, those are the issues that motivated us to redesign the entity clustering process allowing us to eliminate the need for a quality metric and reiterations. Furthermore, in our Modularity based hierarchical record graph clustering approach (GDWM), the adapted Modularity-based algorithm is a second step that involves a minimal number of iterations to optimize the Modularity metric over a record similarity graph, which does not require threshold setting.

We model each soft cluster coming out of the transitive closure process of the DWM as a complete graph. Then, after pruning, the graph of records is further broken down and clustered by optimizing a graph clustering quality heuristic known as Modularity (Schaeffer, 2007). More formally, we combine the equations in 3.1.5, then reformulate each cluster $Q$ as a separate weighted undirected graph where $G' = \{V', E'\}$. Next, the graph is pruned to include only edges available in the dictionary of matched records created in the pairwise matching step, where the edge weights are the precomputed similarity scores. The goal in this second step is to further break down the soft clusters in a hierarchical manner by optimizing through maximizing a graph clustering metric named Modulairty M using the adapted Louvain's method (Blondel et al., 2008). Modularity is a heuristic often used in graph theory (Newman, 2006) to measure the strength of clusters discovered in a graph. Clusters in a graph are often defined by the density of edges between a set of nodes. Hence, the number of connections or edges between clusters characterized by high Modularity is often very low. Modularity is defined in the equation below as M. It measures the difference between the actual number of edges and an expected number of edges between nodes. An expected number of edges between nodes can be considered a random rewiring of the graph given the same nodes. Optimizing Modularity through maximization in a graph is a hard problem and is often tackled through various ways to reduce the number of comparisons between all nodes in a graph.

$$M = \frac{1}{2n} \sum_{i,j} \left[ A_{i,j} - \frac{k_i k_j}{2n} \right] \partial \left( Q_i, Q_j \right)$$

In the equation above, n is the number of edges in the graph, and k is the degree of a node. In addition $A_{i,j}$ Is a weighted adjacency matrix constructed from $E'$. In addition, i and j are indices for each unique record in the file, represented as a vertex in the graph as $\{ v' \in V' \}$. And $\{ e' \in E' \ where \ E' \subseteq V' \times V' \}$ and $e' = A_{i,j}$.

Louvain's method relies on Modularity optimization by efficiently measuring the difference in Modularities between different configurations of node clustering instead of rewiring the network using a simple equation and procedure (Blondel et al., 2008). The algorithm starts by randomly assigning each node in the graph to a unique random cluster. Next, a random node is picked and placed in the same cluster as its closest neighbor. Then Modularity difference is measured between when the node was in its cluster and when the node is in the current neighbors' cluster. If the change in Modularity was positive, then the node is assigned to its new cluster with its neighboring node. If the delta Modularity is negative, the algorithm iterates and places the node in each cluster in the graph until a positive or zero deltas is achieved. If the delta Modularity was zero, the node remains in its cluster. The second phase consolidates the discovered clusters into a new graph of nodes representing whole clusters summing up all involved edge weights. Then the process reiterates on the newly reconstructed graph until no delta Modularity appears. The final clusters from this step can be canonicalized directly. This process is described formally in Algorithm 3.

---

**Algorithm 3:** Adapted Louvain

**Input :** $G = (V', E')$ : graph of nodes their initial clusters and pruned weighted edges
**Result:** $V'$ : each node is assigned to the newly discovered clusters
**while** *True* **do**
    $M_{initial} \leftarrow ComputeModularity(V', E')$
    **for** $v1 \in V'$ **do**
        **for** $v2 \in V'$ **do**
            *update cluster id of v1 to match cluster id of v2*
            $\Delta M \leftarrow$ *compute difference in Modularity M according to Louvain's equation*
            $L_M \leftarrow (v2, \Delta M)$
        **end**
        $c_i \leftarrow \arg\max(L_M)$ *find the largest cluster with the largest delta Modularity* $\Delta M$
        $v2 \leftarrow v1$ *assign the current node to the maximum cluster*
        *updateV'*
    **end**
    $M_{current} \leftarrow ComputeModularity(V', E')$
    **if** $M_{current} \geq M_{initial}$ **then**
        $G \leftarrow$ *whole clusters become new nodes* $V'_{reconstructed}$,
        *edges are summation of weights between clusters* $E'_{reconstructed}$
    **end**
    **else**
        **return** $V'$
    **end**
**end**

---

### 3.1.6 Canonicalization

The clusters are defined and persisted to the disk in this final step through a link index file. The link index file describes the final entity clusters of the unsupervised entity resolution framework as a list of ordered pairs where the first element of each pair represents the cluster identifier the record is assigned to. The cluster identifier is simply the least record identifier in the cluster. The second element of each pair represents the unique record identifier.

## 4 Experiments

### 4.1 Dataset

Table 1. An excerpt from the synthetic dataset used in the evaluation.

| ID | Reference Body |
|---|---|
| A985464 | LLOYD AARON DEAN 2475 SPICEWOOD DR WINSTON SALEM NC 27106 |
| A956423 | LLOYD AARON DEAN 2475 SPICEWOOD DR WINSTON SALEM NC 27106 456 18 2098 |
| A956296 | LLOYD AARON D 2475 SPICEWOOD DR WINSTON SALEM NC 27106 456 18 2098 |
| A830349 | LLOYD AARON DEAN 2475 SPICEWOOD DR WINSTON SALEM NC 27106 456182098 |
| A875214 | ANDREEW AARON STEPHEN 2475 SPICEWOOD DR WINSTON SALEM NC 27106 601 70 6106 |
| A824917 | ANDREW AARON STEPHEN 2475 SPICEWOOD DR WINSTON SALEM NC 27106 601 70 6106 |

Our graph-based clustering method is tested on a synthetic benchmark dataset. We use the synthetic dataset described in (Talburt et al., 2009), a simulator-based data generator that uses probabilistic approaches to generate coherent individual data for persons that do not exist except for S3, which represents generated addresses and names of restaurants that do not exist. The data fields are names, addresses, social security numbers, credit card numbers, and phone numbers mixed in several layout configurations. Some samples are labeled as mixed layout, meaning that each row might come with a different order of those attributes and might not be delimited. The standard label means that all the rows in the data file have the same order and attributes. The generator described in (Ye & Talburt, 2019) used a probabilistic error model to inject various errors in the previously developed simulated dataset. For example, in this excerpt of a generated data file shown in table 1, the first four records are almost identical except for that record A956296 has a missing last name and the format of the phone numbers or whether they exist at all, all are errors injected and generated on purpose. In addition, the last two records are almost identical except for the first name where an intentional error was introduced. A ground truth set recording the actual clusters of the simulated records is then sampled from the generated synthetic database. Next, the corresponding references are pulled from the generated synthetic database to create various sample files with different sizes, levels of quality, and layouts. Sizes of files can vary from 50 to 20K rows, as shown in table 2.

Table 2. A good quality sample indicates that the sample had a limited injected number of errors, while a poor quality sample indicates that a high number of errors were injected. Moreover, a moderate quality indicates a balanced dataset. Below is a table describing the statistics of the dataset we are using.

| Sample Name | Quality | Layout | Number of rows |
|---|---|---|---|
| S1 | Good | Standard | 50 |
| S2 | Good | Standard | 100 |
| S3 | Moderate | Standard | 868 |
| S4 | Good | Standard | 1912 |
| S5 | Good | Standard | 3004 |
| S6 | Moderate | Standard | 19998 |
| S7 | Good | Mixed | 2912 |
| S8 | Poor | Standard | 1000 |
| S9 | Poor | Standard | 1000 |
| S10 | Poor | Mixed | 2000 |
| S11 | Poor | Mixed | 3999 |
| S12 | Poor | Mixed | 6000 |
| S13 | Good | Mixed | 2000 |
| S14 | Good | Mixed | 5000 |
| S15 | Good | Mixed | 10000 |
| S16 | Poor | Mixed | 2000 |
| S17 | Poor | Mixed | 5000 |
| S18 | Poor | Mixed | 10000 |

### 4.2 Experimental setup

Here we compare our hierarchical graph-based record clustering approach (GDWM) built using several modules from the DWM with the same version of the original implementation of the DWM. In our setting, we set the blocking frequency beta β to 6, and the stop word threshold sigma σ is set to 7 for all runs on all samples. Note that these values are fixed and were set based on experience with running both algorithms. They were picked by observing the execution time of each algorithm and the final F1 score, which will be discussed later. When running both algorithms on each of the 18 samples, we varied mu between 0.1 and 0.9. That gives nine runs per sample. When running the DWM, we varied the mu and the epsilon parameters between 0.1 and 0.9. That gives us 81 runs per sample. Both the DWM and our GDWM are implemented in Python with the help of various libraries and packages. We ran the two setups on an Intel i7 Windows machine with 32 GBs of RAM. The code for both setups is available through the link provided at the end of the paper.

We measured the precision and recall against the generated ground truth entity clusters as far as evaluation. The ground truth is a list of each record and its membership cluster identifier. After canonicalization, the saved link index is grouped by the least record identifier in each cluster. Thus, all records belonging to the same cluster will have the same record identifier as the first element in the link index pair. We then loop on each pair in the canonicalized link index and examine whether they belong together in the ground truth. We measure the following statistics against the ground truth for each sample run.

Table 3. Evaluation metrics and statistics

| Statistic Name | Symbol | Description |
|---|---|---|
| True Positives | TP | Number of record pairs that appeared together in the same cluster correctly |
| True Negatives | TN | Number of record pairs that did not appear in the same cluster correctly |
| False Positives | FP | Number of record pairs that appeared together in the same cluster falsely |
| False Negatives | FN | Number of record pairs that did not appear in the same cluster falsely |
| Precision | P | TP / (TP + FP) |
| Recall | R | TP / (TP + FN) |
| F1-Score | F1 | 2 x P . R / P + R |
| Balanced Accuracy | A | TP(TN + FP) + TN(TP + FN) / (TP + FN)(TN + FP) |

### 4.3 Results

Here we present the results after running the 18 samples on different similarity threshold mu μ; as mentioned before that the DWM ran for each mu level and each epsilon level nine times. In addition, the blocking frequency beta and the stop word frequency sigma are both fixed and set to beta=6 and sigma=7. We then chose the epsilon value that yielded the best F1 score and lowest execution time in case of a tie.

Table 4. Average precision, recall, F1 scores, balanced accuracy, and execution time for each algorithm

| Method | Precision | Recall | F1-Score | Balanced Accuracy | Time (seconds) |
|---|---|---|---|---|---|
| DWM [Talburt et al., 2020] | 0.7243 | 0.5782 | 0.6297 | 0.7689 | 203.1 |
| GDWM [In this paper] | **0.8478** | **0.6862** | **0.71479** | **0.8431** | **24.8** |

For the GDWM, we ran on each sample 9 times, varying the similarity threshold mu from 0.1 to 0.9. We recorded both runtime statistics. For each sample run for each algorithm, we chose the lowest mu that yielded the highest F1-score, followed by the lowest execution time in case of a tie. We then averaged precision, recall, F1 scores, balanced accuracies, and execution times for both algorithms across the 18 samples. Table 4 demonstrates the results for each metric and statistic averaged across all 18 samples. For precision, the improvement appears to be the largest as our approach targets an increased precision relying on the fact that the first step of discovering large soft clusters using the graph-based transitive closure approach CC-MR plays the role of a staging step for the Modularity optimization algorithm through stabilizing the recall even at high mu levels. The next step, which involves Modularity optimization on each soft cluster using Louvain's method, is responsible for the increased precision. That increased precision can be attributed to first modeling each set of the soft clustered records into a complete graph then weighting the graph using the previously computed similarity scores in the pairwise matching step. That approach automatically prunes the complete graph since not all the records have precomputed similarity scores. That approach also reduces the Modularity optimization done using Louvain and prevents any possible below zero final modularities that often happen when using Louvain's methods, also known as the resolution limit problem (Blondel et al., 2008). Second, Louvain's method is highly efficient when applied to large graphs due to its hierarchical approach in discovering clusters, and it fits nicely as a second step after transitive closure using CC-MR to provide a fully hierarchical graph clustering method. The hierarchical clustering approach also helps preserve the high recall of the entity clusters across the two steps. That is also shown in the increased balanced accuracy, which is a valuable measure for problems such as entity resolution that is usually characterized by having a class imbalance as the number of matched pairs are usually way less than the number of unmatched pairs causing a very high number of true negatives (Mower, 2005). The significant speed-up achieved across all samples can also be attributed to using the delta Modularity equation provided in the original Louvain's method paper during optimization (Blondel et al., 2008).

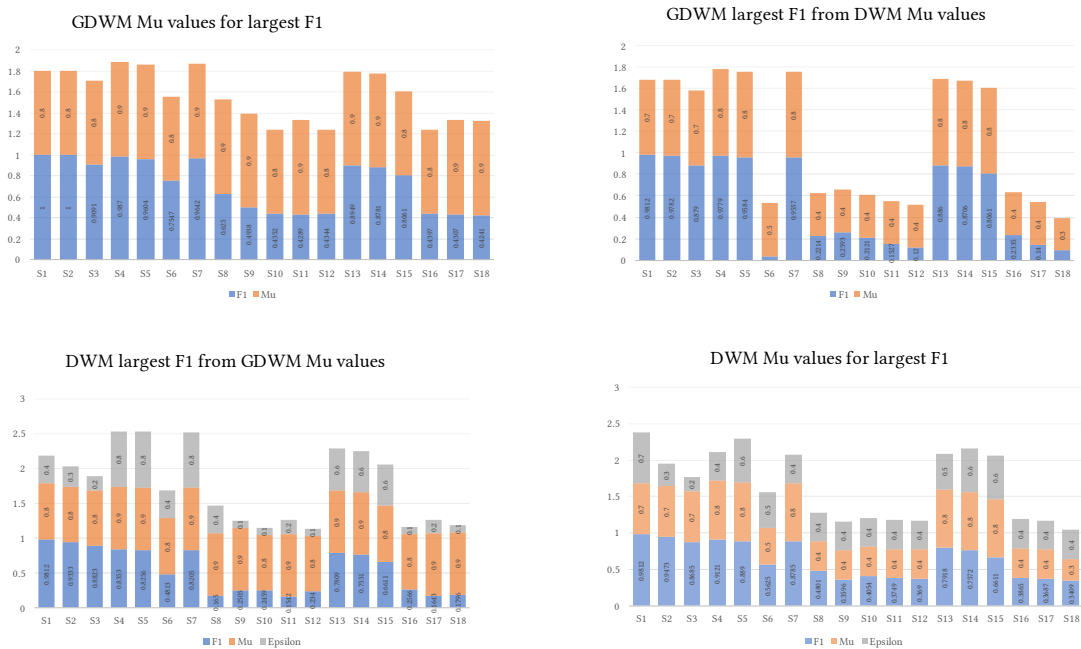

Figure 5. Best F1 score for each sample with its corresponding Mu level for each algorithm

Figure 5 shows the interplay between mu, sample size, and F1 scores across the 18 sample runs. It is worth mentioning that tuning epsilon with the similarity threshold mu can be difficult when running the original implementation of the DWM. Nevertheless, on a sample level and as shown here in figure 5, our approach, the GDWM, has consistently outperformed the iterative approach used in the DWM, especially at mu levels where GDWM had the most significant overall F1 score. Those mu levels where GDWM outperformed DWM are also higher than the mu levels where DWM had the most significant F1 score on its own. However, at those levels, GDWM outperformance was not as consistent as with the original GDWM mu levels. For example, at mu level 0.3 on poor quality sample files such as S8, S9, S10, S11, S12, S16, S17, and S18, DWM slightly outperformed GDWM. That might be attributed to the existence of a mechanism to iterate and adjust quality levels in DWM has made it perform better than GDWM in the case of poor data quality files at low levels of mu.

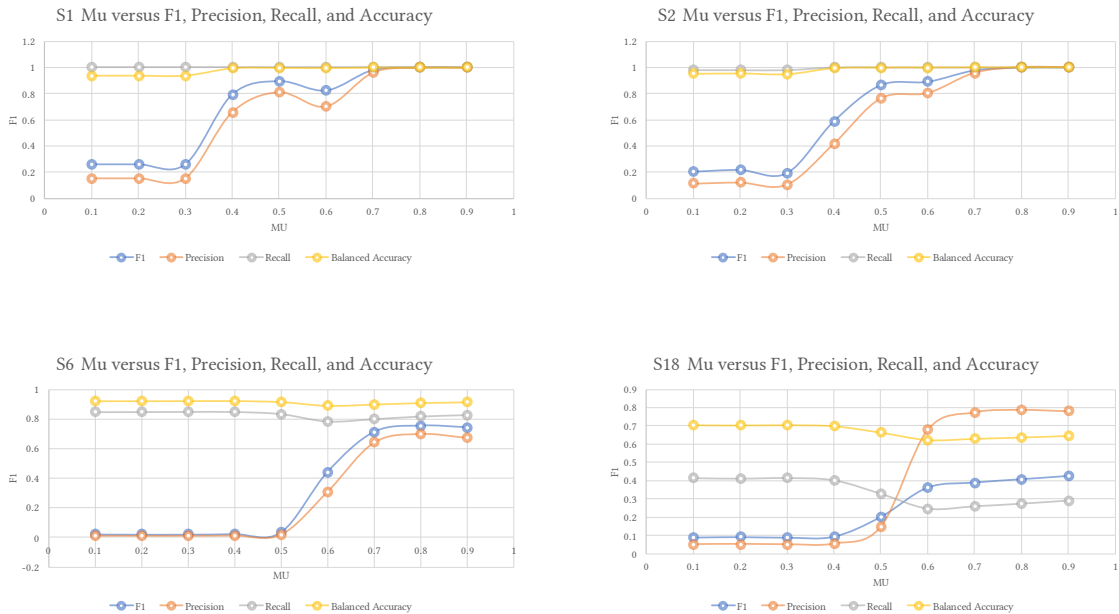

Figure 6. Parameter sensitivity on four samples

From the perspective of parameter sensitivity, at lower mu levels, recall appears to be more sensitive to changes than in higher mu levels. That can be seen in figure 6, showing 4 sample files with different qualities, sizes, and layouts. The figure demonstrates the sensitivity of the similarity threshold mu with the F1 score, precision, recall, and balanced accuracy. An observation is that in small data samples with good quality like S1 and S2, Mu increases precision. Recall seems to be pretty stable across all Mu values because it is more sensitive to blocking frequency beta and block sizes in general, while here, we fixed beta at 6 six and sigma at 7. In large data samples such as S6 with 20K rows, there seems to be an onset for mu where mu before 0.5 does not affect the quality of the clusters produced by GDWM. While low-quality samples like S18 recall starts and continue to be very low at the blocking frequency we set, affecting the overall performance and F1 scores.

## 5 Discussion

Our GDWM approach eliminates the need to reiterate over a big chunk of the remaining records that have not been well clustered, as is the case in the DWM. Our approach also eliminates the need for using a metric such as Shannon's entropy that has been utilized in the DWM to evaluate clustering quality. Instead, it uses a well-known Modularity optimization method for graphs through an efficient implementation of Louvain that does not need threshold setting using the Python programming language. Instead, we rely on the Python library NetworkX (Hagberg et al., 2008). Combining the graph-based transitive closure approach named CC-MR that has been used in the original implementation of the DWM which acts as a first step in identifying connected components or soft clusters by treating the identified matched pairs of records as an edge list of a graph with a Modularity based clustering approach that we introduced here through the implementation of Louvain's method, which acted as a second step in identifying and breaking down the original soft clusters, helped in quickly and early on breaking down any soft clusters that existed right after transitive closure hierarchically and efficiently.

Our approach, however, is not free of limitations despite its efficacy in identifying latent unique entity clusters. For example, our experiments are highly controlled and applied to a synthetic dataset on data for non-existent individuals; despite the usefulness of the dataset, it is not diverse enough to explore the interplay between other parameters when applying our approach or any other unsupervised entity resolution approach for that matter.

Therefore, in the future, we aim to diversify the benchmark datasets used in experimentation. In addition, we plan to study the sensitivity of other crucial parameters like the blocking frequency beta and the stop word frequency sigma instead of fixing them in experiments. In addition, in the future, we aim at studying how many hierarchical levels of clusters are discovered using our approach and how that can inform the discovery of latent entity clusters using our graph method with fewer optimization iterations.

## 6 Conclusion

This paper demonstrated an innovative graph-based Modularity based hierarchical record clustering approach to unsupervised entity resolution named GDWM. Our method combines a graph-based transitive closure algorithm with an adapted Modularity-based graph clustering approach based on Louvain's method in community detection in large networks to provide a two-step, efficient hierarchical record clustering capable of producing exact F1 scores of 100% in some cases. We built on the work done by (Talburt et al., 2020), where they presented a probabilistic self-assessing and iterative approach to unsupervised entity resolution named the Data Washing Machine (DWM). We integrate our GDWM with the DWM to overcome limitations, especially in self-assessment and reiterations where our approach eliminated those needs. In addition, we performed various experiments on 18 synthetic benchmark datasets with different sizes, qualities, and layouts to evaluate our adapted algorithm. Finally, we demonstrated the efficacy and advantages our clustering algorithm provides in terms of F1 scores, accuracy, and speed up of execution time compared with the original implementation of the DWM. We will focus on testing our algorithm on more benchmark datasets in the future. In addition, We will explore ways to make our graph-based hierarchical approach even more efficient as far as execution time and precision.

### Code availability

The code of our graph-based algorithm (GDWM), the synthetic dataset used in the evaluation, in addition to the source code of the Data Washing, Machine are all available as a downloadable code repository at https://bitbucket.org/oysterer/dwm-graph/src/master/GDWM18/. If you have issues downloading the code, please contact the corresponding author at iaebeid@ualr.edu.

### Acknowledgment

This material is based upon work supported by the National Science Foundation under Award No. OIA-1946391.